\documentclass[12pt]{article}

\font\grande=cmr9.5 scaled \magstep4
\font\medio=cmr9.5 scaled \magstep2
\outer\def\beginsection#1\par{\medbreak\bigskip
      \message{#1}\leftline{\bf#1}\nobreak\medskip
\vskip-\parskip
      \noindent}
\usepackage{graphicx} 
\begin{document}
\bibliographystyle {unsrt}

\titlepage

\begin{flushright}
CERN-PH-TH/2014-075
\end{flushright}

\vspace{10mm}
\begin{center}
{\grande Violation of consistency relations}\\
\vspace{0.5cm}
{\grande and the protoinflationary transition}\\
\vspace{1.5cm}
 Massimo Giovannini
 \footnote{Electronic address: massimo.giovannini@cern.ch}\\
\vspace{1cm}
{{\sl Department of Physics, 
Theory Division, CERN, 1211 Geneva 23, Switzerland }}\\
\vspace{0.5cm}
{{\sl INFN, Section of Milan-Bicocca, 20126 Milan, Italy}}
\vspace*{0.5cm}
\end{center}

\vskip 0.5cm
\centerline{\medio  Abstract}
If we posit the validity of the consistency relations, the tensor 
spectral index and the relative amplitude of the scalar and tensor power spectra are both fixed by a single slow roll parameter.  
The physics of the protoinflationary transition can break explicitly the consistency relations causing a reduction of the inflationary 
curvature scale in comparison with the conventional lore. 
After a critical scrutiny, we argue that the inflationary curvature scale, the 
total number of inflationary efolds and, ultimately, the excursion of the inflaton across its 
Planckian boundary are all characterized by a computable theoretical error. 
While these considerations ease some of the tensions between the Bicep2 data and the other satellite observations, 
they also demand an improved understanding of the protoinflationary transition whose physical features may be assessed, in the future,  through 
a complete analysis of the spectral properties of the B mode autocorrelations.
\vskip 0.5cm

\noindent

\vspace{5mm}

\vfill
\newpage
\renewcommand{\theequation}{1.\arabic{equation}}
\setcounter{equation}{0}
\section{Introduction}
\label{sec1}
Inflation must have a limited duration since it cannot extend indefinitely in the past.
The lack of past geodesic completeness of a quasi-de Sitter stage of expansion suggests 
that the initial phase of inflation can be plausibly divided into a preinflationary phase where the background geometry 
decelerates\footnote{The  scale factor of a Friedmann-Robertson-Walker line element shall be denoted by $a$;  the overdot denotes 
a derivation with respect to the cosmic time coordinate.} (i.e. $\dot{a} >0$ but $\ddot{a}<0$) followed by the protoinflationary epoch 
of expansion when $\ddot{a}$ changes its sign. These periods of evolution are likely to be driven by an irrotational fluid. 
The quantum theory of the fluctuations of gravitating, irrotational and relativistic fluids \cite{lukash} has been developed even prior to the actual formulation of the conventional inflationary paradigm and in the context of the pioneering analyses of the relativistic theory of large-scale inhomogeneities \cite{lif}.

The consistency relations are essential for the determination of the inflationary curvature scale as well as for the typical value
of the inflationary potential \cite{wein} at horizon crossing. It has been recently argued that the consistency relations can be violated 
by the protoinflationary physics \cite{mg1}:  a protoinflationary phase containing 
gravitons and fluid phonons can impact differently on the tensor to scalar ratio. In this framework, the consistency relations, even if heuristically assumed in most of the experimental analyses, are not mandatory.

Mixed initial states modify the temperature and polarization anisotropies at large scales and this idea has been scrutinized along various perspectives not only in \cite{mg1} but in a number of previous discussions \cite{two,three,four,five,six,seven,eight}. Temperature-dependent phase transitions may lead to an initial thermal state for the metric perturbations. Second-order correlation effects of the scalar and tensor fluctuations of the geometry can be used to explore the statistical properties of the initial mixed quantum state \cite{six} (see also \cite{nine}). We shall discuss the case of spatially flat models since this is the situation 
suggested by the observational data. The ideas reported here can be however generalized 
to spatially curved background geometries. 

In what follows we are going to examine the violation of the consistency relations in the light of the recent Bicep2 data \cite{bicep2} 
that are seemingly in tension with other satellite observations \cite{WMAP9,planck}. The 
purpose here is not to endorse a model or a mechanism but rather to point out, through specific examples, a set of physical situations 
that are logically plausible.
The consistency relations and their implications for the determination of the inflationary scales are discussed in section \ref{sec2}. In section \ref{sec3} 
the violation of the consistency relations is examined when the protoinflationary phase contains a mixture 
of thermal gravitons. In section \ref{sec4} we consider the more general situations where the initial state contains both thermal gravitons and thermal 
phonons. We finally discuss the third possibility where only the phonons are in a mixed state. In section \ref{sec5} we collect the concluding remarks and 
draw some general lessons.

\renewcommand{\theequation}{1.\arabic{equation}}
\setcounter{equation}{0}
\section{Consistency relations}
\label{sec2}
In the conventional lore, the inflationary curvature and energy scales can be explicitly 
determined provided the tensor to scalar ratio $r_{T}$ is fixed,
 for instance, by the analysis of the B mode angular power spectra. 
Denoting by ${\mathcal A}_{{\mathcal R}}$ and ${\mathcal A}_{T}$
the amplitudes of the scalar and tensor fluctuations 
of the geometry at the conventional pivot\footnote{The choices for the pivot wavenumber are conventional but the common ones 
are $0.002\,\, \mathrm{Mpc}^{-1}$ and $0.05\,\, \mathrm{Mpc}^{-1}$.} wavenumber $k_{p}$,
\begin{equation}
P_{{\mathcal R}}(k_{p}) = {\mathcal A}_{{\mathcal R}}, \qquad P_{T}(k_{p}) = {\mathcal A}_{T},
\label{SP}
\end{equation}
the tensor to scalar ratio $r_{T}$, the tensor spectral index $n_{T}$ and the slow roll parameter $\epsilon$ obey the following 
chain of equalities:
\begin{equation}
r_{T} = \frac{{\mathcal A}_{T}}{{\mathcal A}_{{\mathcal R}}} = 16\epsilon = - 8 n_{T},
\label{first}
\end{equation}
where $\epsilon = - \dot{H}/H^2$ is the slow-roll parameter.
The subscript of the scalar power spectrum refers to the curvature perturbations on comoving orthogonal hypersurfaces  (conventionally 
denoted by ${\mathcal R}$); this is the gauge-invariant variable 
customarily employed when presenting and analyzing observational data since the first WMAP data release \cite{WMAP9} (see also \cite{bard1} for some 
classic references on the Bardeen formalism).

Equation (\ref{first}) defines, in a nutshell, the physical content of the consistency 
relations stipulating that the slow roll parameter  determines simultaneously 
the slope of the tensor power spectrum and the tensor to scalar ratio. From Eq. (\ref{first})  we can determine 
$\epsilon$ if  the value of $r_{T}$ is fixed by observations at $k_{p}$. 

Equation (\ref{first}) holds in the case of single field inflationary models which are the ones conventionally 
confronted with the temperature 
and polarization anisotropies of the Cosmic Microwave Background (CMB in what follows). 
The reason for focussing on single-field models is not only dictated by simplicity 
but also by the absence of any signal of non-gaussianity in spite of all the efforts spent so far 
to justify and discover large non-gaussian signals in the CMB observables. The first equality appearing in Eq. (\ref{first}) rests on the assumption 
that the inflaton fluctuations are the only source of scalar inhomogeneities throughout the development of the inflationary phase. Also 
this second assumption seems well justified in the light of the observed value of $r_{T}$. Since $\epsilon \ll 1$ the scalar fluctuations 
ascribable to a source different from the inflaton $\varphi$ can be, at most, ${\mathcal O}(0.1)$. 

Using Eq. (\ref{first}) the curvature scale of inflation and the typical scale of the inflationary potential are determined as
\begin{equation}
\frac{H}{M_{P}} = \frac{\sqrt{\pi\, {\mathcal A}_{{\mathcal R}} \, r_{T}}}{4}, \qquad \frac{W}{M_{P}^4} = \frac{ 3\, r_{T}\, {\mathcal A}_{{\mathcal R}}}{128}.
\label{second}
\end{equation}
Finally, taking the fourth root of the second relation of Eq. (\ref{second}) and using the definition of the number of efolds the following pair 
of relations can be obtained:
\begin{equation}
E = \biggl(\frac{ 3\, r_{T}\, {\mathcal A}_{{\mathcal R}}}{128}\biggr)^{1/4}, \qquad \biggl|\frac{\Delta \varphi}{\Delta N} \biggr|= \overline{M}_{P} \sqrt{\frac{r_{T}}{8}},
\label{third}
\end{equation}
where $E$ is the typical energy scale of inflation and $|\Delta \varphi/\Delta N|$ denotes the excursion of the inflaton field $\varphi$ with 
the number of efolds $N$. Note that $1/\overline{M}_{P} = \sqrt{8\pi}/M_{P}$ and $M_{P} = 1.22\times 10^{19}$ GeV; both $M_{P}$ and $\overline{M}_{P}$ 
shall be employed hereunder for convenience.

The Bicep2 experiment \cite{bicep2} has observed a $B$-mode polarization of the CMB
that can be well fit by the standard $\Lambda$CDM scenario supplemented by tensors 
with $r_{T}=0.2^{+0.07}_{-0.05}$; the value $r_{T}=0$
is disfavored at more than  $5\sigma$.  It is possible that the actual 
primordial component of $r_{T}$ will be slightly smaller than $0.2$ 
even if  various tests were performed on the data to eliminate systematic effects 
and other contaminations from galactic synchrotron and from polarized 
dust emissions\footnote{After subtraction of some purported foregrounds the values of $r_{T}$ may get closer to the Planck limits \cite{planck} and 
imply $r_{T} = 0.16^{+0.06}_{-0.05}$. It seems too soon to get to definite conclusions on this issue and, therefore, we shall
prefer to set $r_{T}=0.2$ as a fiducial value for the tensor to scalar ratio. Slightly different values of $r_{T}$ will not have any impact for the 
considerations discussed here.}\cite{synch}. 

Using Eqs. (\ref{second}) and (\ref{third}) together with numerical values of $r_{T}$ and ${\mathcal A}_{{\mathcal R}}$ the various scales 
can be written in more explicit terms and they are 
\begin{eqnarray}
\biggl(\frac{H}{M_{P}}\biggr) &=& 9.70\times 10^{-6} \,\biggl(\frac{r_{T}}{0.2}\biggr)^{1/2} \biggl(\frac{{\mathcal A}_{{\mathcal R}}}{2.4\times 10^{-9}}\biggr)^{1/2},
\label{fourth}\\
\biggl(\frac{W}{M_{P}^4}\biggr) &=& 1.12 \times 10^{-11} \, \biggl(\frac{r_{T}}{0.2}\biggr) \biggl(\frac{{\mathcal A}_{{\mathcal R}}}{2.4\times 10^{-9}}\biggr),
\label{fifth}\\
\biggl(\frac{E}{\mathrm{GeV}}\biggr) &=& 2.23 \times 10^{16} \, \biggl(\frac{r_{T}}{0.2}\biggr)^{1/4} \biggl(\frac{{\mathcal A}_{{\mathcal R}}}{2.4\times 10^{-9}}\biggr)^{1/4},
\label{sixth}\\
\biggl|\frac{\Delta\varphi}{\Delta N}\biggr |&=& 3.1 \times 10^{-2}\, \biggl(\frac{r_{T}}{0.2} \biggr)^{1/2} \, M_{P}.
\label{seventh}
\end{eqnarray}
Note, incidentally, that for $\Delta N\simeq 5$,  Eq. (\ref{seventh}) would imply $\Delta\varphi \simeq {\mathcal O}(M_{P})$. According to some viewpoints, the excursion of the inflaton when the relevant scale exit the horizon is crucial to judge of the validity of an effective field theory approach to inflation. 

The maximal number of inflationary efolds 
accessible to large-scale CMB measurements can be derived by demanding that the inflationary event 
horizon redshifted at the present epoch coincides with the Hubble radius today:
\begin{equation}
e^{N_{\mathrm{max}}} = \frac{(2 \pi\, \Omega_{R 0} \,{\mathcal A}_{{\mathcal R}}\,r_{T})^{1/4}}{4}\, \biggl(\frac{M_{P}}{H_{0}}\biggr)^{1/2}\, \biggl(\frac{H}{H_{r}} \biggr)^{1/2- \gamma},
\label{eightha}
\end{equation}
where $\Omega_{R 0}$ is the present energy density of radiation in critical units and $H_{0}^{-1}$ is the Hubble radius today. Equation (\ref{eightha}) is a consequence of Eq. (\ref{first}) since 
 $\epsilon$ must be expressed in terms of $r_{T}$. For the pivotal set of parameters of Eqs. (\ref{fourth})--(\ref{sixth}),  Eq. (\ref{eightha}) becomes:
 \begin{eqnarray}
 N_{\mathrm{max}} &=& 61.49 + \frac{1}{4} \ln{\biggl(\frac{h_{0}^2 \Omega_{R 0}}{4.15 \times 10^{-5}} \biggr)} - \ln{\biggl(\frac{h_{0}}{0.7}\biggr)}
 \nonumber\\
 &+& \frac{1}{4} \ln{\biggl(\frac{{\mathcal A}_{{\mathcal R}}}{2.4 \times 10^{-9}}\biggr)} + \frac{1}{4} \ln{\biggl(\frac{r_{T}}{0.2}\biggr)} + \biggl(\frac{1}{2} - \gamma\biggr) 
 \ln\biggl{(\frac{H}{H_{r}}\biggr)}.
 \label{eighthb}
\end{eqnarray} 
In Eqs. (\ref{eightha}) and (\ref{eighthb})  $\gamma$ accounts for the possibility of a delayed reheating terminating at a putative scale $H_{r}$
smaller than the Hubble rate during inflation; $\gamma$ controls the expansion rate in the intermediate phase. Since the reheating scale cannot be smaller than the one of nucleosynthesis, 
 $H_{r}$ can be as low as $10^{-44} M_{\mathrm{P}}$ (but not smaller) corresponding to a reheating scale occurring just prior to the formation of the light nuclei.
 If $\gamma - 1/2 >0$ (as it happens if $\gamma = 2/3$ when the post-inflationary background is dominated by dust \cite{LL}), $N_{\mathrm{max}}$ diminishes in comparison with the sudden reheating (i.e. $H=H_{r}$) and $N_{\mathrm{max}}$ can become ${\mathcal O}(47)$.
Conversely if $\gamma - 1/2 <0$ (as it happens in $\gamma = 1/3$ 
when the post-inflationary background is dominated by stiff sources \cite{LL,qi}), $N_{\mathrm{max}}$ increases.  Finally, if $H_{r} = H$ (or, which is the same, if $\gamma=1/2$) there is a sudden transition 
between the inflationary regime and the post-inflationary epoch dominated by radiation. In spite of its dependence on ${\mathcal A}_{{\mathcal R}}$ and $r_{T}$, the value of $N_{\mathrm{max}}$ has then a theoretical error. Based on the previous considerations and on the maximal excursion of $\gamma$ we can write
\begin{equation}
N_{\mathrm{max}} = 61.49 \pm  14.96.
\label{eighthc}
\end{equation}
 If the total number of inflationary efolds $N_{\mathrm{t}}$ 
is larger than $N_{\mathrm{max}}$ (i.e. $N_{\mathrm{t}} > N_{\mathrm{max}}$), then the redshifted value of the inflationary event horizon 
is larger than the present value of the Hubble radius.

Let us finally mention that when $N_{t} \gg N_{\mathrm{max}}$ we expect, at least in conventional inflationary 
models, that any finite portion of the Universe gradually loses the memory of an initially imposed anisotropy or inhomogeneity so that the Universe attains the observed regularity regardless of the initial boundary conditions. The previous statement expresses, in practical terms, what is often dubbed cosmic no-hair 
conjecture (see e.g. \cite{NH0} and references therein). 
This viewpoint has been questioned long ago by Barrow in a specific class of power-law inflationary backgrounds \cite{NH1}.
In general terms it is difficult to claim when the no-hair conjecture is valid. For instance in some classes of anisotropic inflationary models where the gauge fields 
are coupled to the inflaton, magnetic hairs can persist in spite of the number of efolds \cite{NH0,NH2}. In 
the conventional framework adopted here we can expect that the no-hair conjecture holds, as it can be established by carefully analyzing the problem 
within the gradient expansion \cite{NH3}. This is therefore a relevant demand when assigning 
the protoinflationary inhomogeneities. 
 
\renewcommand{\theequation}{3.\arabic{equation}}
\setcounter{equation}{0}
\section{Violation of the consistency relations}
\label{sec3}
The results of Eqs. (\ref{fourth})--(\ref{seventh}) depend on the assumed 
consistency between the scalar and tensor modes of the geometry. This consistency can be violated even in the context 
of single field inflationary models by the protoinflationary dynamics. 
Suppose, for sake of simplicity, that during the protoinflationary background the phonons and the gravitons are not in the 
vacuum but in a mixed state \cite{mg1,three} (see also \cite{five,six,seven,eight}).

Assuming either thermal or kinetic equilibrium, the initial state is
described by a density matrix with Bose-Einstein distribution i.e. for each species 
\begin{equation}
\hat{\rho} = \sum_{\{n\}} P_{\{n \}} |\{n \}\rangle \langle \{n \}|,
\qquad P_{\{n\}} = \prod_{\vec{k}} \frac{\overline{n}_{k}^{n_{k}}}{( 1 + \overline{n}_{k})^{n_{k} + 1 }},
\label{twelve}
\end{equation}
where $\overline{n}_{k}$ is the average multiplicity of each Fourier mode and $ |\{n \}\rangle = |n_{\vec{k}_{1}} \rangle \otimes  |n_{\vec{k}_{2}} \rangle \otimes  |n_{\vec{k}_{3}} \rangle...$. The ellipses stand for all the occupied modes of the field. In particular the average multiplicities of gravitons and phonons are: 
\begin{equation}
\overline{n}^{\mathrm{ph}}_{k} = \frac{1}{e^{\omega^{\mathrm{ph}}/k_{\mathrm{T}}}-1}, \qquad \overline{n}^{\mathrm{gr}}_{k} = \frac{1}{e^{\omega^{\mathrm{gr}}/k_{\mathrm{T}}} -1},
\label{thirteen}
\end{equation}
where, in units $\hbar= c = k_{B} =1$,  
\begin{equation}
\omega^{\mathrm{ph}}= k \, c_{\mathrm{s}}, \qquad \omega^{\mathrm{gr}}= k,\qquad k_{T} =T.
\label{wn}
\end{equation}
We shall posit that phonons and gravitons have the same temperature but this assumption can be dropped if only approximate (kinetic) equilibrium holds between 
the different species\footnote{In practice we shall either assume that 
the phonons and gravitons have the same temperature (as it would follow from considerations of local thermal equlibrium) or different temperatures 
(as it would follow in the case of kinetic equilibrium).}. 

The density matrix of the scalar and tensor fluctuations of the geometry is the direct product of the mixed quantum states of the phonons and of the gravitons
 $\hat{\rho} = \hat{\rho}_{\mathrm{phonons}}\, \otimes\, \hat{\rho}_{\mathrm{gravitons}}$; the power spectra are obtained by tracing the product of the 
density matrix with the relevant field operators: 
\begin{eqnarray}
&& \mathrm{Tr}\biggl[ \hat{\rho}\,\hat{{\mathcal R}}(\vec{x}, \tau)\, \hat{{\mathcal R}}(\vec{x} + \vec{r}, \tau) \biggr] = 
\int \frac{d k}{k} {\mathcal P}_{{\mathcal R}}(k,\tau) \frac{\sin{k r}}{k r},
 \label{fourteen}\\
 && \mathrm{Tr}\biggl[ \hat{\rho}\,\hat{h}_{ij}(\vec{x}, \tau) \,\hat{h}^{ij}(\vec{x} + \vec{r}, \tau) \biggr] = 
 \int \frac{d k}{k} {\mathcal P}_{{\mathrm{T}}}(k,\tau) \frac{\sin{k r}}{k r},
\label{fifteen}
\end{eqnarray}
where $\hat{R}(\vec{x},\tau)$ and $\hat{h}_{ij}(\vec{x},\tau)$ (with $\hat{h}_{i}^{i} =0 =\nabla_{i}\, \hat{h}^{i}_{j}$) denote the field 
operators corresponding, respectively, to the scalar and tensor modes of the geometry; $\tau$ is the conformal time coordinate. If during the protoinflationary phase there are both thermal gravitons and thermal phonons the final power\footnote{The two mode functions 
for the two polarizations of the graviton coincide; moreover, the initial fluid phonons are the normal modes of the 
protoinflationary fluid that can be quantized in terms of the Lukash variable \cite{lukash} which is nothing but 
the curvature perturbation on comoving orthogonal hypersurfaces and has been derived in the case of an irrotational fluid. }
spectra are \cite{mg1}
\begin{eqnarray} 
{\mathcal P}_{{\mathcal R}}(k,\tau) &=&   \frac{8}{3 M_{\mathrm{P}}^4} \biggl(\frac{W}{\epsilon}\biggr) \biggl(\frac{k}{k_{p}}\biggr)^{n_{\mathrm{s}} -1}( 2 \overline{n}^{\mathrm{ph}}_{k} +1),
\label{nineth}\\
{\mathcal P}_{{\mathrm{T}}}(k,\tau) &=&   \frac{128}{3 } \biggl(\frac{W}{M_{\mathrm{P}}^4}\biggr) \biggl(\frac{k}{k_{p}}\biggr)^{n_{\mathrm{T}}} ( 2 \overline{n}^{\mathrm{gr}}_{k} +1), 
\label{tenth}
\end{eqnarray}
where the scalar and tensor spectral indices are, respectively, $n_{\mathrm{s}} = 1 - 6 \epsilon + 2 \overline{\eta}$ and $n_{\mathrm{T}} = - 2 \epsilon$,
with $\overline{\eta}= \overline{M}_{P}^2\,W_{,\varphi\varphi}/W$. From the point of view of the Schr\"odinger description the quantum states involved 
in this problem are squeezed thermal states \cite{three} (see also \cite{sqt}).

Let us first consider, for simplicity, the situation where the phonons are absent (i.e.  $\overline{n}_{\mathrm{ph}} =0$ ) while the gravitons are in a thermal state.
From the ratio of the tensor and scalar power spectra of Eqs. (\ref{nineth}) and (\ref{tenth}) we obtain, at the pivot scale $k_{p}$ the following relation between $\epsilon(k_{p})$ and $r_{T}(k_{p})$:
\begin{equation}
\epsilon(k_{p}) = \frac{r_{T}(k_{p})}{16} \tanh{\biggl(\frac{k_{p}}{2 \, k_{T}}\biggr)}.
\label{sixteenth}
\end{equation}
The explicit value of $k_{T}$ at the present time depends on the maximal temperature of the protoinflationary epoch:
\begin{eqnarray}
k_{T} &=& T_{\mathrm{max}} \, Q\, e^{- N_{t}} \biggl(\frac{2 \Omega_{R0}}{\pi {\mathcal A}_{{\mathcal R}} \epsilon} \biggr)^{1/4} \, \biggl(\frac{H_{r}}{H}\biggr)^{\gamma - 1/2} 
\sqrt{\frac{H_{0}}{M_{P}}},
\nonumber\\
T_{\mathrm{max}} &=&  \biggl(\frac{45}{4 \pi^3 g} \biggr)^{1/4} \, \sqrt{H \, M_{P}},
\label{seventeenth}
\end{eqnarray}
and $Q$ is the fraction of $T_{\mathrm{max}}$ ascribable to the gravitons. Furthermore, the backreaction constraints during the 
protoinflationary stage demand that  $Q<1$. The value of $k_{T}$ can be 
expressed in units of the present value of the Hubble rate:
\begin{equation}
\frac{k_{T}}{H_{0}} = \biggl(\frac{45\, \Omega_{R0}}{2 \pi^3 g} \biggr)^{1/4} \, Q\, \sqrt{\frac{M_{P}}{H_{0}}}  \, \biggl(\frac{H_{r}}{H}\biggr)^{\gamma -1/2} \, e^{- N_{t}}.
\label{eighteen}
\end{equation}
Equation (\ref{eighteen}) implies an even more stringent relation between $k_{T}$, the total number of efolds $N_{t}$ and the critical number of efolds 
$N_{\mathrm{crit}}(k)$:
\begin{equation}
\frac{k}{2 k_{T}}= \frac{g^{1/4}}{Q} \, e^{N_{t} - N_{\mathrm{crit}}(k)},
\label{nineteenth}
\end{equation}
where $N_{\mathrm{crit}}(k)$ is defined as:
\begin{equation}
N_{\mathrm{crit}}(k) = 66.25 - \ln{ \biggl(\frac{k}{0.002\mathrm{Mpc}^{-1}}\biggr)}  + \biggl(\gamma -\frac{1}{2} \biggr)\ln{\biggl(\frac{H_{r}}{H}\biggr)} 
+ \frac{1}{4} 
\ln{\biggl(\frac{h_{0}^2 \Omega_{R0}}{4.15 \times 10^{-5}}\biggr)}.
\label{twenty}
\end{equation}

Equation (\ref{sixteenth}) implies that $16 \epsilon(k_{p}) \neq r_{T}(k_{p})$.  The consistency relations are then violated and,
as it follows from Eq. (\ref{nineteenth}), $k_{p}/ k_{T}$ is a function of  $N_{t} -N_{\mathrm{crit}}(k_{p})$. The function parametrizing the relation 
between $\epsilon$ and $r_{T}$ can be written as 
${\mathcal F}^2(N_{t}-N_{\mathrm{crit}}, g, Q)$ where 
\begin{equation}
 {\mathcal F}(x, g, Q) = \sqrt{\tanh{\biggl(\frac{g^{1/4}}{Q} e^{x}\biggr)}}.
\label{twentythird}
\end{equation}
Note that $N_{t}$ is unknown while $N_{\mathrm{crit}}(k_{p})$ is affected by a theoretical error comparable 
to the one of Eqs. (\ref{eighthb}) and (\ref{eighthc}) and coming from the different post-inflationary histories. Both numbers 
can be fixed by plausible theoretical guesses but they are phenomenologically undetermined.
Thus, the inflationary scales previously introduced can be expressed as follows:
\begin{eqnarray}
\biggl(\frac{H}{M_{P}}\biggr) &=&  9.70\times 10^{-6} \,\biggl(\frac{r_{T}}{0.2}\biggr)^{1/2} \biggl(\frac{{\mathcal A}_{{\mathcal R}}}{2.4\times 10^{-9}}\biggr)^{1/2}\,\, {\mathcal F}(N_{t}-N_{\mathrm{crit}}, g, Q),
\label{twentyfirst}\\
\biggl(\frac{W}{M_{P}^4}\biggr) &=&  1.12 \times 10^{-11} \, \biggl(\frac{r_{T}}{0.2}\biggr) \biggl(\frac{{\mathcal A}_{{\mathcal R}}}{2.4\times 10^{-9}}\biggr)\,\,{\mathcal F}^2(N_{t}-N_{\mathrm{crit}}, g, Q),
\label{twentysecond}\\
\biggl|\frac{\Delta \varphi}{\Delta N}\biggr| &=& 3.1 \times 10^{-2}\, \biggl(\frac{r_{T}}{0.2} \biggr)^{1/2} {\mathcal F}(N_{t}-N_{\mathrm{crit}}, g, Q)\,\,
 M_{P},
\label{twentyfourth}
\end{eqnarray} 
where $r_{T}\equiv r_{T}(k_{p})$ and $N_{\mathrm{crit}}\equiv N_{\mathrm{crit}}(k_{p})$. Note that $N_{\mathrm{crit}}(k_{p}) = 66.25$ in the sudden reheating approximation (i.e. $\gamma = 1/2$ and $H= H_{r}$).  If we assume that the reheating 
is not instantaneous but delayed by a long post-inflationary phase stiffer than radiation we shall have that 
Eqs. (\ref{twentyfirst}) and (\ref{twentysecond}) hold with a different $N_{\mathrm{crit}}(k)$. In practice we shall consider the same excursion 
of $15$ efolds already discussed in the case of $N_{\mathrm{max}}$ (see Eq. (\ref{eighthc})) and bear in mind that $N_{\mathrm{crit}}(k_{p}) \simeq 66.25\pm 15$. 

The properties of  $ {\mathcal F}(x, g, Q) $ depend mildly on $g$ and $Q$ and more crucially on $N_{t}$ and $N_{\mathrm{crit}}$. Let 
us therefore choose $g=2$ for the two polarizations of the graviton and $Q=0.1$ and let us assume
the sudden reheating approximation (i.e. $N_{\mathrm{crit}}=66.25$ for the fiducial set of parameters). Thus for $N_{t} = 50$ we have, from 
Eqs. (\ref{twentyfirst}), (\ref{twentysecond}) and (\ref{twentyfourth}) that: 
\begin{equation}
\biggl(\frac{H}{M_{P}} \biggr)= 9.90 \times 10^{-9},\qquad \biggl(\frac{W}{M_{P}^4}\biggr)=1.16\times 10^{-17}, \qquad \biggl|\frac{\Delta \varphi}{\Delta N}\biggr| = 3.16\times 10^{-5}\, M_{P}.
\label{twentyfifth}
\end{equation}
Let us finally consider a far more extreme situation, namely the case where the reheating is delayed down to the nucleosynthesis scale 
and the expansion is stiffer than radiation (i.e. for instance $\gamma=1/3$ in Eq. (\ref{twenty})). Then Eq. (\ref{twentyfifth}) becomes
\begin{equation}
\biggl(\frac{H}{M_{P}}\biggr) = 5.47 \times 10^{-12},\qquad \biggl(\frac{W}{M_{P}^4}\biggr)=3.57\times 10^{-24}, \qquad \biggl|\frac{\Delta \varphi}{\Delta N}\biggr| = 1.75 \times 10^{-8}\, M_{P},
\label{twentyfifthb}
\end{equation}
always for a total number of efolds $N_{t} =50$. 

Let us remark, incidentally, that an excursion of the inflaton of $\Delta N \simeq 5$ or even $10$ does not hit severely the Planckian boundary. 
The variation of $Q$ corresponds to a lower temperature of the gravitons in units of $T_{\mathrm{max}}$. If $Q$ diminishes, for instance, by two orders of magnitude (from $0.1$ to $10^{-3}$) the overall 
effect on the curvature scale corresponds to one order of magnitude (i.e. $10^{-8} \to 10^{-7}$ in Eq. (\ref{twentyfifth}) and 
$10^{-12} \to 10^{-11}$ in Eq. (\ref{twentyfifthb})). 

The total number of efolds is usually considered, for practical purposes, between $50$ and $60$. For instance 
the Planck collaboration \cite{planck} gives for the number of efolds a possible excursion between $50$ and $60$.
A growth in the total number of efolds increases the inflationary scales; for instance 
when  $N_{t}= 60$ and in the sudden reheating approximation (i.e. $N_{\mathrm{crit}} = 66.25$)
we shall  have that $H/M_{P}= 1.46\times 10^{-6}$, $W/M_{P}^{4} = 2.5\times 10^{-13}$ and $|\Delta\varphi/\Delta N| = 4.69\times 10^{-3}$. These 
figures are a bit smaller than (but of the same order of)  the ones given in the previous section.  

In the limit $N \gg N_{\mathrm{crit}}$ the consistency relations are recovered since, as it can be explicitly checked, 
\begin{equation}
\lim_{N_{t} \gg N_{\mathrm{crit}}}  {\mathcal F}(N_{t}-N_{\mathrm{crit}}, g, Q) \to 1. 
\label{ped15}
\end{equation}
Equation (\ref{ped15}) agrees with the no-hair conjecture \cite{NH0,NH1,NH3} and it is consistent with the whole 
approach. Arbitrary modifications of the initial state violating the no-hair conjecture 
may lead to misleading conclusions unless the features of the model allow for 
such a violation as speculated in the past \cite{NH0} and also more recently 
\cite{NH0,NH2} in different frameworks. Unfortunately neither $N_{\mathrm{crit}}$ nor $N_{t}$ are fixed (or even bounded) by the 
no-hair conjecture. 

The relevance of the initial conditions of large-scale fluctuations in the determination of the excursion of the scalar field have been reported in \cite{nine} 
in an implicit model suggesting that the effective theory can be saved if the initial state is a mixed state. We agree 
with the idea of \cite{nine} insofar as the initial mixed states may break the consistency relations since this 
is the suggestion already discussed in \cite{mg1} (see also \cite{three}). We do not agree, however, with the absence 
of a critical number of efolds following, on  a general ground, from the validity of the no-hair conjecture 
in the conventional set-up of single field inflationary models. We finally disagree with the statement that 
the presence of a mixed state in the initial conditions is sufficient to guarantee a strong violation of the consistency relations (see, in particular, the discussion
of the following section).

It would be tempting to identify $N_{\mathrm{crit}}$ with $N_{\mathrm{max}}$ since numerically the two quantities are roughly 
coincident. However, at the risk of being pedantic it is appropriate to remark that since Eqs. (\ref{twentyfirst}) and (\ref{twentysecond}) depend on the number of efolds the determination 
of $N_{\mathrm{max}}$ is more involved than in the case when the consistency relations are valid. 
In particular the relation that determines $N_{\mathrm{max}}$ is given, in this case, by 
\begin{equation}
e^{N} \, \biggl(\frac{H}{M_{P}}\biggr)^{\gamma -1}_{N} = \biggl( 2 \Omega_{R0}\biggr)^{1/4} \biggl(\frac{H_{r}}{M_{P}}\biggr)^{1-\gamma} \biggl(\frac{M_{P}}{H_{0}} \biggr)^{1/2},
\label{ped1}
\end{equation}
where the subscript at the left hand side reminds that $(H/M_{P})$ depends, this time, on the number of efolds. Equation (\ref{ped1}) is not an algebraic 
equation but it can be solved in three different limits (i.e. $N\gg N_{\mathrm{crit}}$, $N\ll N_{\mathrm{crit}}$ and $N= {\mathcal O}(N_{\mathrm{crit}})$). Notice, finally, that when $\gamma=1/2$ and the consistency relations are restored Eq. (\ref{ped1}) gives exactly Eq. (\ref{eightha}).

\renewcommand{\theequation}{4.\arabic{equation}}
\setcounter{equation}{0}
\section{Thermal phonons and thermal gravitons}
\label{sec4}
The results obtained in the previous section suggest that an initially mixed state leads to a violation 
of the consistency relations provided the total number of efolds does not exceed the critical number of efolds. 
We shall now address a slightly different question and ask if the presence of a mixed 
state during the protoinflationary stage is also sufficient to guarantee 
a violation of the consistency relations. 

There is an amount of fine-tuning in assuming that only the 
gravitons are thermal while the phonons are not.  Reversing the argument we could say that the amount of breaking 
of the consistency relations reflects our ignorance on the total duration of the inflationary phase but also 
some sort of postulated asymmetry in the initial conditions of the large-scale flucutaions. While it may well be that this is exactly what the observational 
data demend it is nonetheless interesting to relax this assumption.

Thus, if thermal gravitons and thermal 
phonons are simultaneously present all the considerations developed in the previous section can be repeated 
with few main differences. The relation between $\epsilon(k_{p})$ and $r_{T}(k_{p})$ (and the consequent breaking of the consistency 
relations) is different from the one of Eq. (\ref{sixteenth}) and it is given by 
\begin{equation}
\epsilon(k_{p}) = \frac{r_{T}(k_{p})}{16} \frac{\tanh{\biggl(\frac{k_{p}}{2 \, k_{T_{g}}}\biggr)}}{\tanh{\biggl(\frac{c_{s}\,k_{p}}{2 \, k_{T_{ph}}}\biggr)}}, 
\label{ped2}
\end{equation}
where $c_{s}$, as already mentioned, is the sound speed of the phonons. In Eq. (\ref{ped2}) we allow for different thermal wavelengths 
of the phonons and of the gravitons.  
Unlike section \ref{sec3} we have that $g \geq 3$ (since also the phonons should be counted as thermal species); finally 
Eq. (\ref{twentythird}) gets modified as follows:
\begin{equation}
 \overline{{\mathcal F}}(x, \,g, \,Q,\,c_{s}) =  \frac{{\mathcal F}(x, g, Q)}{{\mathcal F}(x, g, Q/c_{s})}.
 \label{ped3}
\end{equation}
The form of Eq. (\ref{ped3}) suggests 
a much less important effect on the inflationary curvature and energy scales.

More specifically, the analog of Eqs. (\ref{twentyfifth}) and (\ref{twentyfifthb}) does seem to depend on $N_{\mathrm{crit}}$ and it is given by 
\begin{equation}
\frac{H}{M_{P}} = 1.25 \times 10^{-5},\qquad \frac{W}{M_{P}^4}=1.93\times 10^{-11}, \qquad \biggl|\frac{\Delta \varphi}{\Delta N}\biggr| = 4.07\times 10^{-2},
\label{ped4}
\end{equation}
for $N_{t}=50$, $g =3$ (the two polarizations of the graviton plus the phonon) and $c_{s} = 1/\sqrt{3}$ (i.e. in the case of a preinflationary phase 
dominated by radiation).  

The rationale for the previous finding stems from the $\overline{{\mathcal F}}(x, \,g, \,Q,\,c_{s})$ 
that implies a weak breaking of the consistency relations. While for $N_{t} \gg N_{\mathrm{crit}}$ the consistency relations are recovered, when 
$N_{t} < N_{\mathrm{crit}}$ we can roughly approximate 
\begin{equation}
r_{T}(k_{p}) \simeq 16 \, c_{s}\, \epsilon(k_{p}) \biggl(\frac{T_{g}}{T_{ph}}\biggr). 
\label{ped5}
\end{equation}
In thermal equilibrium $T_{g} \simeq T_{ph}$ and the breaking is therefore proportional to the sound speed $c_{s}$.

As suggested in \cite{mg1} it is tempting to speculate that future measurements of the tensor spectral index from the slope of the B mode polarization can decide if and how the consistency relations are broken by the initial conditions. In spite of this we must also admit 
that by relaxing the tuning of the initial conditions and by allowing for the presence of thermal phonons 
the violation of the consistency relations becomes progressively less relevant.

A complementary way of addressing this issue is through the running of the spectral index. The running of the spectral index implied by this type of models 
has the same qualitative features for the scalar and for the tensor modes of the geometry with the difference that 
the scalar spectral index is directly measured while the tensor spectral index can only be inferred from the 
consistency relations, if valid. In particular, defining the ratio $\kappa = k_{p}/(2 k_{T})$  we have 
\begin{eqnarray}
n_{s} &=& 1  - 6 \epsilon + 2 \overline{\eta} + q_{s} +\frac{1}{2}\alpha_{s} \ln{(k/k_{p})},
\label{ped6}\\
n_{T} &=& - 2 \epsilon +q_{T}  + \frac{1}{2}\alpha_{T} \ln{(k/k_{p})},
\label{ped7}
\end{eqnarray}
where, in the case discussed here, the running parameters are
\begin{eqnarray}
\alpha_{s} &=& \frac{4 c_{s} \kappa}{\sinh{2 c_{s} \kappa}}\,  \biggl( -1 + \frac{2 c_{s} \kappa }{\tanh{2  c_{s} \kappa}} \biggr),
\label{ped8}\\  
\alpha_{T} &=& \frac{4  \kappa}{\sinh{2  \kappa}}\,  \biggl( -1 + \frac{2 \kappa }{\tanh{2  \kappa}} \biggr),
\label{ped9}
\end{eqnarray} 
while $q_{s}$ and $q_{T}$ are given by 
\begin{equation}
q_{s} = - \frac{c_{s} \kappa}{\cosh{c_{s} \kappa} \,\sinh{c_{s}\kappa}}, \qquad 
q_{T} = - \frac{\kappa}{\cosh{ \kappa} \,\sinh{\kappa}};
\label{ped10}
\end{equation}
$q_{s}$ and $q_{T}$ go both to $0$ for $\kappa \gg 1$ and to $-1$ for $\kappa\ll 1$; $\alpha_{s}$ and $\alpha_{T}$ 
are both positive. The running of the scalar spectral index does not seem to relax the compatibility 
between the Bicep2 data and the Planck upper bound on the tensor to scalar ratio, since, naively, the running 
is always positive (instead of negative) and, furthermore, the constant contribution changes the spectral 
slope too radically at large scales especially in the case of the scalar modes. It is however not clear, at the moment, if this conclusion applies also in the 
present case where the consistency relations are violated since the determination 
of the running of the spectral index from the fits assumes the validity of the consistency relations.

Let us mention, for sake of completeness that we have, in principle, a third final possibility 
stipulating that the breaking of the consistency relations instead of being concentrated in the graviton 
sector, or equally shared between phonons and gravitons is rather due to thermal phonons alone. 
In this third case the analog of Eqs. (\ref{twentyfirst}), (\ref{twentysecond}) and (\ref{twentyfourth})
\begin{eqnarray}
\biggl(\frac{H}{M_{P}}\biggr) &=&  9.70\times 10^{-6} \,\biggl(\frac{r_{T}}{0.2}\biggr)^{1/2} \biggl(\frac{{\mathcal A}_{{\mathcal R}}}{2.4\times 10^{-9}}\biggr)^{1/2}\,\, {\mathcal F}^{-1}(N_{t}-N_{\mathrm{crit}}, g, Q/c_{s}),
\label{ped11}\\
\biggl(\frac{W}{M_{P}^4}\biggr) &=&  1.12 \times 10^{-11} \, \biggl(\frac{r_{T}}{0.2}\biggr) \biggl(\frac{{\mathcal A}_{{\mathcal R}}}{2.4\times 10^{-9}}\biggr)\,\,{\mathcal F}^{-2}(N_{t}-N_{\mathrm{crit}}, g, Q/c_{s}),
\label{ped12}\\
\biggl|\frac{\Delta \varphi}{\Delta N}\biggr| &=& 3.1 \times 10^{-2}\, \biggl(\frac{r_{T}}{0.2} \biggr)^{1/2} {\mathcal F}^{-1}(N_{t}-N_{\mathrm{crit}}, g, Q/c_{s})\,\,
 M_{P}.
\label{ped13}
\end{eqnarray} 
According to Eqs. (\ref{ped11}), (\ref{ped12}) and (\ref{ped13}) the inflationary scales get larger than the 
conventional values. This is due to the fact that the tensor to scalar ratio instead of being determined 
by Eq. (\ref{sixteenth}) is now given by 
\begin{equation}
r_{T}(k_{p}) = 16 \epsilon(k_{p}) {\mathcal F}^{2}(N_{t}-N_{\mathrm{crit}}, g, Q/c_{s}).
\label{ped14}
\end{equation}
In spite of minor differences due to the sound speed we can say that the breaking of the consistency relations 
goes actually in a direction that is opposite to the one to the one suggested by the observational data and, in this 
sense, it is purely academic. 

\renewcommand{\theequation}{5.\arabic{equation}}
\setcounter{equation}{0}
\section{Concluding remarks}
\label{sec5}

During  the protoinflationary transition the consistency relations can be violated 
even in the case of conventional single field models. This possibility entails 
various theoretical uncertainties that may interfere either constructively or destructively.

The total duration of inflation is unknown and it is customarily assigned in terms of the number of efolds, 
i.e. the natural logarithm of the total increase of the scale factor during inflation.
According to the no-hair conjecture, when inflation lasts beyond some critical number of efolds any finite portion of the Universe 
gradually loses the memory of an initially imposed anisotropy or inhomogeneity. Since the preinflationary phase is likely 
to be dominated by radiation it is plausible that the normal modes of the geometry and of the sources 
will be in a mixed rather than in a pure state. The simplest possibility 
along this line of thinking, is the one where phonons and gravitons obey a Bose-Einstein 
distribution. The critical value of inflationary efolds is then determined by 
the temperature of the initial mixed state. When the total number of efolds greatly exceeds the critical value, according 
to the no-hair conjecture, the Universe attains the observed regularity regardless of the initial boundary conditions.

Three extreme physical situations can be envisaged. In the first case 
 only the gravitons are in a mixed state. In the second case phonons and gravitons are both in a thermal state, possibly with 
 different temperatures while in the third case only the phonons are in a mixed state. 
The violation of the consistency relations depends on the 
asymmetry of the initial data: if only the gravitons are thermal the inflationary scales can be safely lowered depending on the 
total number of efolds and on the critical number of efolds.
Conversely the breaking of the consistency relations is moderate when both phonons 
and gravitons are in kinetic equilibrium. This means that the initial mixed state for the cosmological perturbations is not sufficient 
to guarantee a sizable violation of the consistency relations.
If the tension between the Bicep2 data and the other satellite observations will persist the potential violations of the consistency 
relations can offer a unique handle on the nature of the initial data, as the examples reported here suggest.

At the moment the consistency relations do not follow from any empirical evidence but just from 
plausible arguments that can be evaded, as the present considerations demonstrate.
The future observations must then devise direct tests of the consistency relations.
In particular, the forthcoming  programs  will be essential for the accurate determination 
of the tensor spectral index $n_{\mathrm{T}}$ from the slope of the B-mode power spectrum. 
As already suggested in \cite{mg1} it is tempting to speculate 
 that independent measurements of $r_{T}$ (from the B mode amplitude) and of $n_{T}$ (from the B mode slope)  
 may offer novel diagnostics of the role played by primordial phonons and gravitons in setting the initial conditions of 
 large-scale gravitational perturbations.

\end{document}